\documentclass[conference]{IEEEtran}
\IEEEoverridecommandlockouts
\usepackage{cite}
\usepackage{amsmath,amssymb,amsfonts}
\usepackage{algorithmic}
\usepackage{graphicx}
\usepackage{textcomp}
\usepackage{xcolor}
\def\BibTeX{{\rm B\kern-.05em{\sc i\kern-.025em b}\kern-.08em
    T\kern-.1667em\lower.7ex\hbox{E}\kern-.125emX}}

\IEEEoverridecommandlockouts
\makeatletter
\def\ps@IEEEtitlepagestyle{%
  \def\@oddfoot{\mycopyrightnotice}%
  \def\@evenfoot{\mycopyrightnotice}%
}
\newcommand{\mycopyrightnotice}{%
  \begin{minipage}{\textwidth}\footnotesize\centering
    \copyright~2026 IEEE. Personal use of this material is permitted.
    Permission from IEEE must be obtained for all other uses, in any
    current or future media, including reprinting/republishing this
    material for advertising or promotional purposes, creating new
    collective works, for resale or redistribution to servers or lists,
    or reuse of any copyrighted component of this work in other works.
  \end{minipage}}
\makeatother

\begin{document}

\title{Beyond VMAF: Towards Application-Specific Metrics for Teleoperation Video

\thanks{The research was financially
supported by the Federal Ministry of Research, Technology and Space of Germany (BMFTR) within the
project ASUR, No. 03ZU2105BA.}
}
\author{\IEEEauthorblockN{Ines Trautmannsheimer}
\IEEEauthorblockA{\textit{Chair of Automotive Technology} \\
\textit{Technical University of Munich}\\
Munich, Germany \\
ines.trautmannsheimer@tum.de \\
0009-0004-0147-01832}
\and
\IEEEauthorblockN{Richard Grauberger}
\IEEEauthorblockA{\textit{Chair of Automotive Technology} \\
\textit{Technical University of Munich}\\
Munich, Germany \\
richard.grauberger@tum.de}
\and
\IEEEauthorblockN{Frank Diermeyer}
\IEEEauthorblockA{\textit{Chair of Automotive Technology} \\
\textit{Technical University of Munich}\\
Munich, Germany \\
diermeyer@tum.de}

}

\maketitle

\begin{abstract}
Automated driving has made remarkable progress, yet situations still arise where human intervention is necessary. Teleoperation provides a scalable solution to address such cases, enabling remote operators to support vehicles without being physically present. In this context, video transmission forms the operator’s primary source of situational awareness, making video quality a decisive factor for both safety and task performance.
In an online study, participants rated compressed video sequences from the Zenseact Dataset and provided subjective quality ratings. These ratings were then used to retrain the Video Multi-Method Assessment Fusion (VMAF) model, yielding an adapted variant tailored to teleoperation.
The retrained model demonstrated improved alignment with human ratings compared to the original 4K VMAF. In particular, RMSE decreased from 10.36 to 8.83, and MAD from 8.71 to 6.38, corresponding to improvements of 15\% and 27\%, respectively. These results highlight that incorporating domain-specific data can enhance the predictive power of established quality metrics in safety-critical applications.
At the same time, Outlier cases emerged in which videos received high objective scores despite noticeable degradations in regions critical for the driving task. 

\end{abstract}

\section{Introduction}
\begin{figure}[h]

    \includegraphics[width=\linewidth]{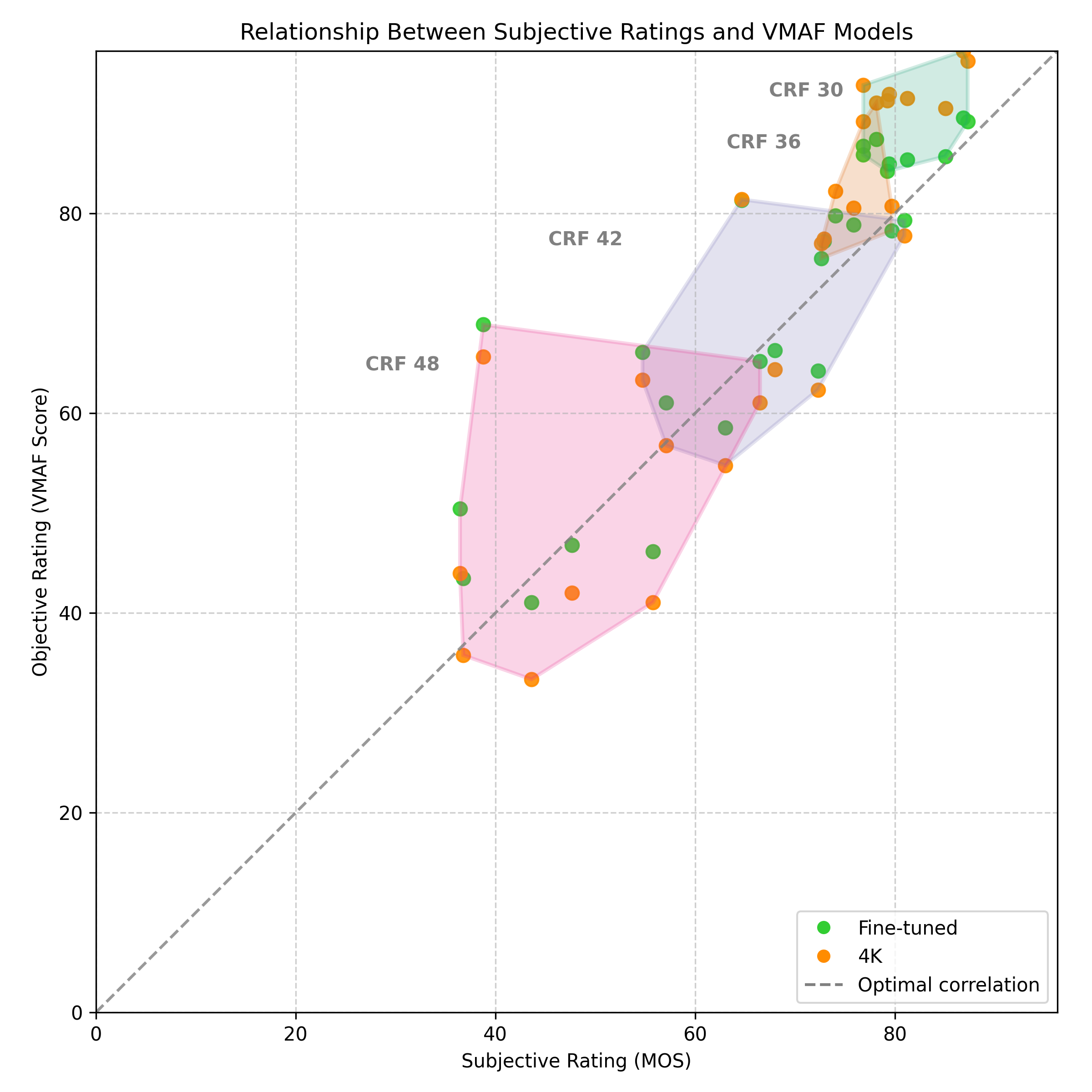}
    \caption{Relationship between subjective human ratings and model predictions. The fine-tuned model captures perceptual differences more accurately in high-quality conditions but both underestimates degradations in low-quality regions.}
    \label{fig:mos vs vmaf}

\end{figure}
Automated driving continues to progress rapidly, with several deployments demonstrating the feasibility of driverless operation in restricted domains \cite{reuthersZoox,Waymo2025DC}. Yet full autonomy is neither universal nor uniformly reliable: long-tail scenarios, adverse weather, and rare edge cases still require human judgement. A practical solution is teleoperation, which involves remote support or remote driving on demand \cite{brechtEvaluationTeleoperationConcepts2024}.

In teleoperation, the operator’s primary situational awareness is mediated by video. Consequently, the quality of video as perceived by humans becomes a safety-relevant factor. Latency, compression artifacts, and sudden network degradation can impair hazard anticipation, distance estimation, and motor control. 
A compression artifact or localized blur in a critical region, such as a pedestrian crossing or oncoming traffic, can directly affect safety. 
Yet conventional video quality assessment (VQA) metrics such as PSNR, (MS-)SSIM, and Netflix’s VMAF \cite{winklerEvolutionVideoQuality2008, wangMultiscaleStructuralSimilarity2003, 
bjontegaard2001calculation, netflix_vmaf} were developed for entertainment streaming, where the goal is perceptual comfort rather than decision accuracy. 
This limitation makes existing VQA unsuitable as a feedback signal for safety-aware teleoperation systems. 
Therefore, we investigate whether a domain-specific adaptation of VMAF can better align with human judgements in teleoperation scenarios and serve as a reliable control signal for adaptive, safety-aware interfaces. 

We design a teleoperation-specific questionnaire that captures perceptual dimensions beyond generic Quality of Experience, namely loss of detail, drivability, and situational awareness, and run an online study on curated driving scenes with controlled compression levels. Using these MOS labels, we retrain the regression component of VMAF to obtain a teleoperation-tuned variant. We then compare against the standard VMAF-4K baseline and analyze failure cases where task-critical regions drive human ratings.

The adapted model improves correlation with human judgements and reduces error versus VMAF-4K on our dataset, with the largest gains at medium compression levels. Qualitative analyses show that remaining mismatches concentrate in scenes where semantic importance is highly localized, motivating region and semantics-aware extensions.

\section{Objective Video Quality Metrics}
\label{sec:objective video quality metrics}
With the rise of large-scale video streaming services, the need for reliable and efficient methods to evaluate compression quality became increasingly evident. In response, numerous objective metrics \cite{xiaoMultiscaleStructureSIMilarity2011, netflix_vmaf, winklerEvolutionVideoQuality2008} were developed or adapted to enable automated assessment, reducing the reliance on human raters and allowing systematic, data-driven optimization of video encoding.

\subsection{Signal Based Metrics}
One of the simplest and earliest objective metrics is the Peak Signal-to-Noise Ratio (PSNR). PSNR measures the ratio between the maximum possible signal power and the power of the distortion (noise) introduced by compression, based on the mean squared error (MSE) between the original and the compressed image.
Due to its simplicity, PSNR remains relevant for codec evaluation and standardization, however, PSNR essentially represents a heuristic of a signal’s noisiness and does not always align with human visual perception. As Winkler and Mohandas \cite{winklerEvolutionVideoQuality2008} demonstrate, PSNR results often deviate from subjective perception, as it constitutes a purely pixel-wise comparison. 

This is one of the reasons why the Structural Similarity Index Measure (SSIM) was developed to obtain a more perceptually accurate assessment. Horé and Ziou \cite{horeImageQualityMetrics2010} evaluated PSNR and SSIM against each other and found that PSNR is more sensitive to Gaussian blur artifacts, while SSIM provides more accurate values for JPEG compression \cite{wallace1992jpeg, wang2001image}. Although SSIM is mathematically simple and already incorporates aspects of the human visual system, it was further extended to a multi-scale metric (MS-SSIM) by Wang et al. \cite{wangMultiscaleStructuralSimilarity2003}. 
The motivation behind this extension was that human perception of image quality depends not only on local structures but also on information across different spatial resolutions. MS-SSIM therefore evaluates luminance, contrast, and structural similarities across several downscaled versions of the same image and aggregates these into a single score. 

Recent studies, however, have shown that the performance gap between SSIM and MS-SSIM is not as clear-cut as originally assumed. By carefully reparametrising SSIM (SS-SSIM), it is possible to achieve comparable or even superior correlations with human ratings \cite{bakurovStructuralSimilarityIndex2022}. 
In particular, data-driven optimization revealed that the structural component of SSIM is more important than luminance or contrast, and that the commonly used default constants ($K_1$, $K_2$), as well as the sliding-window size, should be revised depending on the viewing conditions of the underlying subjective studies. 
These findings suggest that MS-SSIM’s advantage lies less in its multi-scale nature per se but rather in the implicit adaptation to human viewing distances. Properly tuned, single-scale SSIM can therefore perform equally well while remaining computationally less demanding. 

In Wang et al. \cite{wangImageQualityAssessment2004}, the authors present that SSIM evaluates luminance, structure, and contrast equally. However, later studies \cite{horeImageQualityMetrics2010, bakurovStructuralSimilarityIndex2022} show that this does not correspond to human perception and that structural features are most important for human vision. Zheng et al. \cite{zhengVideoQualityAssessment2024} also compare full-reference, reduced-reference, and no-reference metrics and show that MS-SSIM correlates better with MOS but lacks the ability to detect temporal artifacts and complex perceptual masking effects. This underscores the rationale behind the development of learning-based approaches, which aim to capture perceptual relevance and context-aware distortions that extend beyond handcrafted features \cite{zhengVideoQualityAssessment2024}. 
As already mentioned, signal-oriented metrics use equal weightings for the different scales or components of the metric. 
This simplification disregards the fact that the human visual system prioritizes structural information and contextual relationships over simple luminance or contrast differences. 
Learning-based metrics overcome this limitation by learning perceptually optimal feature weightings directly from subjective data. 
As Zhou et al. \cite{zhouPerceptualVisualQuality2025} highlight, frameworks such as  Video Multi-Method Assessment Fusion (VMAF) \cite{LiPracticalPerceptualVideo2017, netflix_vmaf} represent an intermediate stage between analytical formulations (e.g., MS-SSIM) and fully deep perceptual models, combining multiple perceptual indicators through data-driven fusion to more closely align with human judgments.

\subsection{From Analytical Metrics to Perceptual Learning: The Role of VMAF}

The VMAF metric was developed by Netflix to provide a perceptually accurate assessment of video quality across diverse compression levels and viewing conditions \cite{LiPracticalPerceptualVideo2017, netflix_vmaf}, addressing key limitations of signal-based metrics. 
Unlike classical measures such as PSNR or SSIM, which rely on fixed analytical formulations, VMAF integrates multiple perceptual features, such as Visual Information Fidelity, Detail Loss Metric (DLM), and Mean Co-located Pixel Difference into a single composite quality score. 
These features are fused via a supervised learning model based on Support Vector Regression (SVR) \cite{smola2004tutorial}, trained on large-scale subjective datasets to maximize correlation with human MOS. 
By learning the perceptual weighting of structural, contrast, and temporal cues, VMAF bridges the gap between traditional signal-based approaches and data-driven perceptual quality models, providing a reliable proxy for human visual judgment in streaming scenarios.

A major strength of VMAF lies in its open-source implementation, which has made it a de facto standard across both research and industry. 
Topiwala et al. \cite{topiwalaVMAFVariantsUnified2021} replaced the standard SVR with feed-forward neural networks and motion-aware features, achieving up to 9\% higher correlation with subjective ratings and demonstrating the impact of alternative regressors. 
Similarly, Zhang et al. \cite{zhangEnhancingVMAFNew2021} and Chen et al.\cite{chenUnsupervisedCurriculumDomain2021} explored domain-adaptive learning to enhance performance on unseen datasets, while Müller et al. \cite{mullerMachinelearningBasedVMAF2023} retrained VMAF for HDR video using a 3D Convolutional Neural Network (3D-CNN), successfully extending it to scenarios with high luminance and wide color gamut.

These efforts illustrate that VMAF’s predictive power is inherently tied to the domain on which it was trained. 
Models optimized for consumer SDR streaming may not generalize to environments with different visual characteristics, such as high-motion, low-latency, or teleoperation contexts. 
Building on this observation, the present work retrains the regression model within the open-source VMAF framework using a domain-specific dataset. 
This allows the metric to learn feature weightings that reflect the perceptual demands of teleoperated driving, thereby aligning automated quality assessment more closely with human experience in this specialized application domain.

\section{Teleoperation and Video Quality}
\label{sec:teleoperation and video quality}
When developing metrics for evaluating video quality in teleoperation, it is important to recognize the domain-specific requirements, which differ significantly from those of entertainment streaming. It should be noted that although the GUI consists of more information than the video streams, both experts and non-experts consider it to be the most important element \cite{wolfUserCenteredTeleoperationGUI2025}. Hoffmann et al. \cite{hoffmannQuantifyingInfluenceImage2022a} investigated the influence of image quality on operator performance in the presence of dynamic obstacles. Their results revealed significant variations in reaction times and subjective evaluations across different image quality levels. This demonstrates that, unlike in conventional streaming, the evaluation of image quality in teleoperation also has a safety aspect, with performance and accuracy playing a major role. 
In addition to empirical findings on operator performance, Hoffmann and Diermeyer \cite{hoffmannSystemsTheoreticSafetyAssessment2021} conducted a systems-theoretic safety assessment of teleoperated road vehicles, identifying human–machine interaction and perception limitations as critical system hazards. In the context of teleoperation, it is crucial to distinguish between areas that are imperative for the execution of the driving task and those that are of negligible significance and can be largely disregarded. However, in the event that these areas are affected by compression artifacts or substandard quality, the impact on the perceived quality experienced by the operator is greater than in less significant areas \cite{drorOptimizingTrafficSigns2024a, neumeierDataRateReduction2022}. 
Their analysis highlights that degraded visual feedback or delayed video transmission can lead to unsafe control actions, even when other subsystems function correctly. 
This perspective further reinforces that video quality in teleoperation is not merely a question of perceptual comfort but a key element of the overall system safety. 
Accordingly, quality metrics for this domain must be designed to reflect their safety-critical role rather than aesthetic or purely perceptual fidelity.

\section{Data Generation}

\subsection{Choosing a Video-Dataset}
The primary stage in the creation of a newly developed metric is the establishment of a domain-specific dataset. This dataset is utilized for the training of a model and for the purpose of validation. It is imperative that this dataset is made open source. The initial stage of the process is the selection of suitable videos, which are then compressed and evaluated by human operators. The dataset was required to be compatible with the designated research vehicle, which necessitated a minimum size of 1920 x 1200px.
Furthermore, the video had to appear smooth, which is why we excluded datasets below 10~Hz. Furthermore, datasets focusing on lidar or other sensors were excluded, due to the inferior image quality and the presence of too many artifacts in the original. In order to map the domain adequately and mitigate the weaknesses of VMAF, as Neumeier \cite{neumeierVisualQualityTeleoperated2020} discovered with Waymo videos, it was necessary to capture footage at multiple times of day and in a range of weather conditions. The result of our research was the Zenseact dataset \cite{alibeigi2023zenseactopendatasetlargescale}, from which we selected 39 videos with suitable scenarios and shortened them to a length of 8 seconds. This resulted in the identification of three distinct categories: day-good weather, day-bad weather, and night-good weather. It is regrettable that the dataset under consideration did not include any videos of night-bad weather. In addition, despite conducting a comprehensive search of other pertinent datasets, no videos that met the specified criteria could be identified.

\subsection{Creating a Domain Specific Questionnaire}
\label{sec:Creating a Domain Specific Questionnaire}
In order to collect subjective reference data for model training, we developed a domain-specific questionnaire tailored to the characteristics of teleoperated driving. 
As explained in Section~\ref{sec:teleoperation and video quality}, teleoperation requires quality assessments that go beyond purely aesthetic evaluations. Instead, these subjective assessments should focus on functionality and situational understanding, as well as objects that are important for the driving task. 
Despite the fact that ITU-T Recommendation P.910 \cite{ITU_T_P910_2023} provides standardized methods for evaluating video quality, its one-dimensional MOS scale does not adequately capture the perceptual dimensions relevant to representing these requirements.

The development of the questionnaire was a collaborative process involving teleoperation specialists, with the aim of evaluating perceived video quality across four dimensions reflecting the specific Requirements of teleoperation:

\begin{itemize}
    \item Loss of detail: measures the operator's ability to perceive visual features and judge distances and sizes.
    \item Drivability: evaluates how well the visual information supports the driving task, obstacle detection, and path estimation.
    \item Situational awareness: measures the perceived completeness of the scene, the predictability of other traffic participants, and confidence in decision-making.
    \item Reflection after the original video: compares the operator's first impression of the compressed video with the reference of the actual situation to quantify the perceived loss of information and safety relevance.
\end{itemize}

Each dimension was evaluated using several items on a 5-point Likert scale in accordance with ITU conventions. This scale was used to ensure the comparability of the data with existing MOS data while extending the semantic scope to teleoperation-specific constructs and normalizing it to the VMAF scale. 
The last section of the study included a multiple-choice question with objective verification elements, the identification of observed objects, with the aim of verifying the objective perception of the test subjects. 
This multidimensional design allows for both a detailed analysis of perception-related subcomponents and the derivation of composite quality ratings tailored to operator performance and task safety.

\subsection{Online User Study}
The selected videos and data set were then used to create the online study for obtaining human evaluations. 
All videos were encoded using the H.264/AVC codec implemented in FFmpeg’s \texttt{libx264} encoder \cite{tomar2006converting}. 
Encoding was performed with the \texttt{slow} preset and a variable-quality mode based on the Constant Rate Factor (CRF) parameter, ensuring consistent perceptual quality across the video while allowing the bitrate to vary. 
Four compression levels were generated for each video using CRF values of 30, 36, 42, and 48, resulting in 156 unique video pairs comprising the original and its compressed variants. 

The CRF method provides an efficient perceptual trade-off between bitrate and distortion, with lower values (e.g., 18–24) being visually lossless and higher values (e.g., 48–51) introducing strong artifacts \cite{tomar2006converting} \cite{neumeierVisualQualityTeleoperated2020}. 
This configuration follows common practices in subjective video quality studies for teleoperation, where bitrate stability is less critical than maintaining consistent visual quality across frames.

To this end, a dedicated survey website was created using React \cite{react}. This approach enabled the successful implementation of our survey requirements and ensured the smooth execution of the study. To participate, it was necessary to meet specific prerequisites. Firstly, the participant must be in possession of a German driving license. Secondly, the screen must have a minimum size of 25 inches. Thirdly, the participant must have normal or corrected normal vision. The website's design precluded the navigation of users through the browser, whether in a forward or reverse direction, and did not request full-screen view. Initially, the study participants were asked to provide demographic information to ensure the validity of the subsequent data. In order to test the subjects' visual acuity and display settings, a short Landolt ring test \cite{ISO_8596_2017} with size and contrast, and an Ishihara test \cite{ishihara38} were conducted. To execute this process, it was first necessary to calibrate the screen dimensions. This calibration was also utilized in the subsequent section to ensure that the dimensions of the videos were consistent for all subjects. In the experiment, the participants were presented with ten scenarios in random order, with the restriction that each compression level was included at least once and no scenario occurred more than once. Initially, the subjects were presented with a compressed video and asked to answer questions related to the first three dimensions, in order to assess the situation. The participants were then shown the original video and asked questions about the fourth dimension. They were invited to indicate whether and how they would revise their assessment now that all the information was available. To replicate the teleoperation scenario, where the video is live and cannot be replayed, it was necessary to disable replay for the video. 

\section{Training of VMAF}

In this work, the same underlying framework was used to retrain VMAF on data collected from the teleoperation-specific user study described in Section~\ref{sec:Creating a Domain Specific Questionnaire}. 
The objective of the study was to adapt the perceptual model to the visual and functional characteristics of teleoperation, with the aim of improving its correlation with human judgments in this domain. 
To prepare the data, it was necessary to convert all original and compressed video files from MP4 to the YUV420 format using FFmpeg. 
The conversion was performed in 8-bit YUV 4:2:0 planar format (\texttt{YUV420p}), which represents the most widely used chroma subsampling standard in both research and industry.  
This step ensures that the evaluation is not affected by container metadata or codec-specific preprocessing.  
The selection of YUV420 as the primary format is predicated on its prevalence in both research and industrial quality evaluation pipelines. 
 
The training dataset was constructed by mapping each compressed video to its corresponding reference video using unique identifiers, namely content ID and asset ID. 
For each pair, the Differential Mean Opinion Score obtained from the questionnaire was assigned as the training label. 
To circumvent the issue of overfitting, the dataset was meticulously segmented into non-overlapping sets based on the scene. The allocation is 80\% for training and 20\% for validation. 
This approach was adopted to guarantee that the model did not encounter the same scene in both sets, even under different compression levels. 
The categorization of scenes and subsequent distribution are outlined below: 
The experiment was conducted under the following conditions:
\textit{Day – Good Weather} (15 training, 3 validation), 
\textit{Day – Bad Weather} (8 training, 2 validation), 
and \textit{Night – Good Weather} (9 training, 2 validation).

Before the retraining, baseline evaluations were carried out using Netflix's standard VMAF models (HD and 4K) to measure the difference from the human ratings in teleoperation. 
We excluded the mobile model, as the teleoperation use case involves large display setups and the mobile variant has shown lower sensitivity to fine-grained quality differences. 
These baseline results were stored for later comparison with those of the retrained model. 

For retraining, the original SVR structure of VMAF was retained. However, the regression targets were replaced with the MOS-derived teleoperation ratings. 
The new model was trained using the libvmaf framework’s dataset configuration interface to define resolution, YUV format and the association between reference and distorted videos. 
The performance of the model was evaluated on the validation set using Spearman Rank-Order Correlation Coefficient and Root Mean Square Error (RMSE) between predicted and subjective scores. 
%%This retrained model serves as the foundation for the proposed Teleoperation-Focused Multi-Method Assessment Fusion (ToFMAF) metric.

\section{Results}
We evaluated 39 scenes at four compression levels (CRF 30/36/42/48) across three context categories (day–god weather, day–bad weather, night–good weather). Subjective MOS were compared against two objective models: a fine-tuned VMAF variant and the VMAF-4K baseline. The HD model was not investigated further, as the MAD and RMSE were significantly higher than those of the 4K model.%, as can be seen in the figure~\ref{fig:error-comparison}.

% \begin{figure}[t]
%   \centering

%     \includegraphics[width=\linewidth]{figures/comparison_models_bars_validation.png}
%     \caption{Comparison of prediction errors (RMSE and MAD) between the fine-tuned teleoperation-specific model and the baseline VMAF-4K. Both metrics quantify the deviation between subjective MOS and model predictions across compression levels. The adapted model shows consistently lower error values, indicating closer alignment with human perception.}
%     \label{fig:error-comparison}
%     \end{figure}

\begin{figure}[t]
  \centering

    \includegraphics[width=\linewidth]{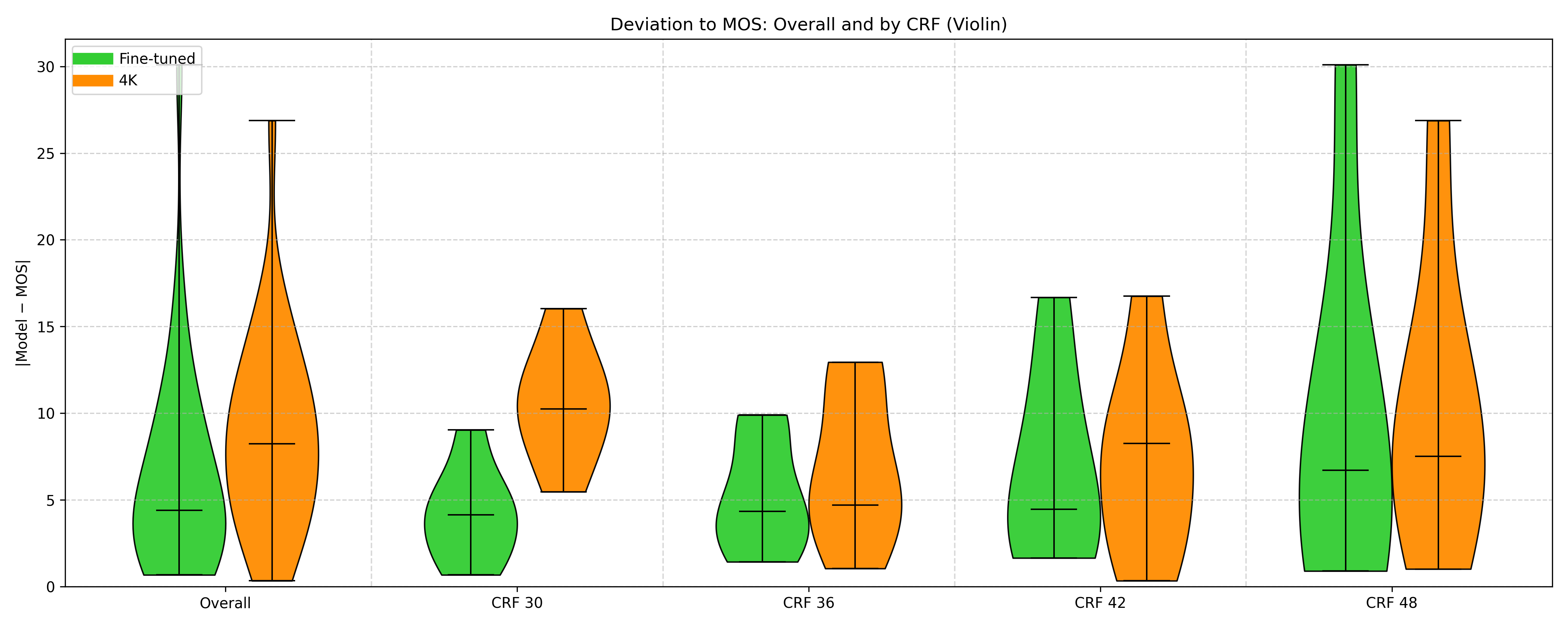}
    \caption{Error distributions of model–human disagreement across compression levels. The fine-tuned variant yields lower and more consistent deviations from human judgments, especially where perceptual quality remains task-relevant. At extreme compression, both models diverge, reflecting the limits of feature-based metrics in capturing task-specific perception in teleoperation.}
    \label{fig:error violins}
\end{figure}

\subsection{Participants}

The online study included a total of 264 participants. A total of 180 people were recruited and paid via an online platform, on the condition that they completed the full questionnaire. The remaining 84 participants came from either the university environment or were incomplete submissions.  
The average age of the participants was 39.75 years ($SD = 12.84$), with a range from 18 to 74 years.  

When it came to driving experience, 96.6\% of those taking part said they had a valid driver's license, with an average experience of 19.48 years ($SD = 12.60$).  
Only 7.95\% of the participants had prior experience in remotely operating a vehicle, which confirms that the majority of the participants were typical end-users rather than trained teleoperators.  
This distribution was intentional. It was to approximate realistic user perception in future teleoperation scenarios. In these scenarios, professional remote drivers will remain a minority.  

A predominance of male participants (68.9\%) was shown in the gender distribution, followed by female (30.7\%) and non-binary (0.4\%) respondents.  

The collective of participants constitutes a sufficiently extensive and demographically variegated sample for reliable statistical analysis of subjective video quality perception in teleoperation contexts.

\subsection{Questionnaire}

\subsubsection{Validity of the Questionnaire}

In order to evaluate the internal validity and consistency of the developed teleoperation-specific questionnaire, inter-item reliability and convergence across scales were analysed.  
All subscales (Detail Loss, Drivability, and Situational Awareness) exhibited strong positive correlations with the aggregated MOS, confirming coherent construct behaviour.  
Cronbach’s $\alpha$ across all items was $\alpha = 0.97$, indicating high internal reliability of the measurement instrument.  
Furthermore, consistent directional effects were observed in responses across scales with constant rate factor (CRF), confirming that participants interpreted the questions as intended and that the questionnaire accurately captured perceived video degradation.  

Across all subjective scales and CRF had a strong and consistent effect on perceived video quality (Kruskal–Wallis, $H = 123.22$, $p < .001$).  
Significance was found for all CRF pairs in paired comparisons (Holm-adjusted Mann–Whitney, all $p < .001$), confirming the convergence of validity across scales.  
Effect sizes (Cliff’s $\delta$) increased monotonically with compression: medium-to-large for 30 vs 36 ($\delta = 0.61$), large for 36 vs 42 ($\delta = 0.76$), and very large for 42 vs 48 ($\delta = 0.92$).  
This monotonic increase across all subscales provides strong evidence that perceptual degradation is reliably measured by the questionnaire as compression intensifies.

\subsubsection{Effects of Compression on Perceived Quality}

To analyze the influence of compression level on perceived quality, the responses were first tested for normality using the Shapiro–Wilk test.  
All CRF-level samples met the normality criterion ($p > .05$), allowing the use of a one-way ANOVA.  
This revealed a highly significant effect of compression on perceived quality ($F(3, 152) = 247.38$, $p < .001$).  

Tukey’s HSD post hoc tests confirmed that each compression step produced statistically distinct mean ratings:  
CRF~30~$>$~36 ($p = .0003$),  
CRF~30~$>$~42 ($p < .001$),  
CRF~30~$>$~48 ($p < .001$),  
CRF~36~$>$~42 ($p < .001$),  
CRF~36~$>$~48 ($p < .001$),  
and CRF~42~$>$~48 ($p < .001$).  

These findings substantiate the hypothesis that participants could reliably differentiate between all four compression levels, even between moderate steps (30~→~36~→~42), demonstrating a high degree of perceptual sensitivity.  
The consistency of this pattern across all questionnaire scales further validates the robustness of the subjective data.  

\begin{figure}[t]
  \centering

    \includegraphics[width=\linewidth]{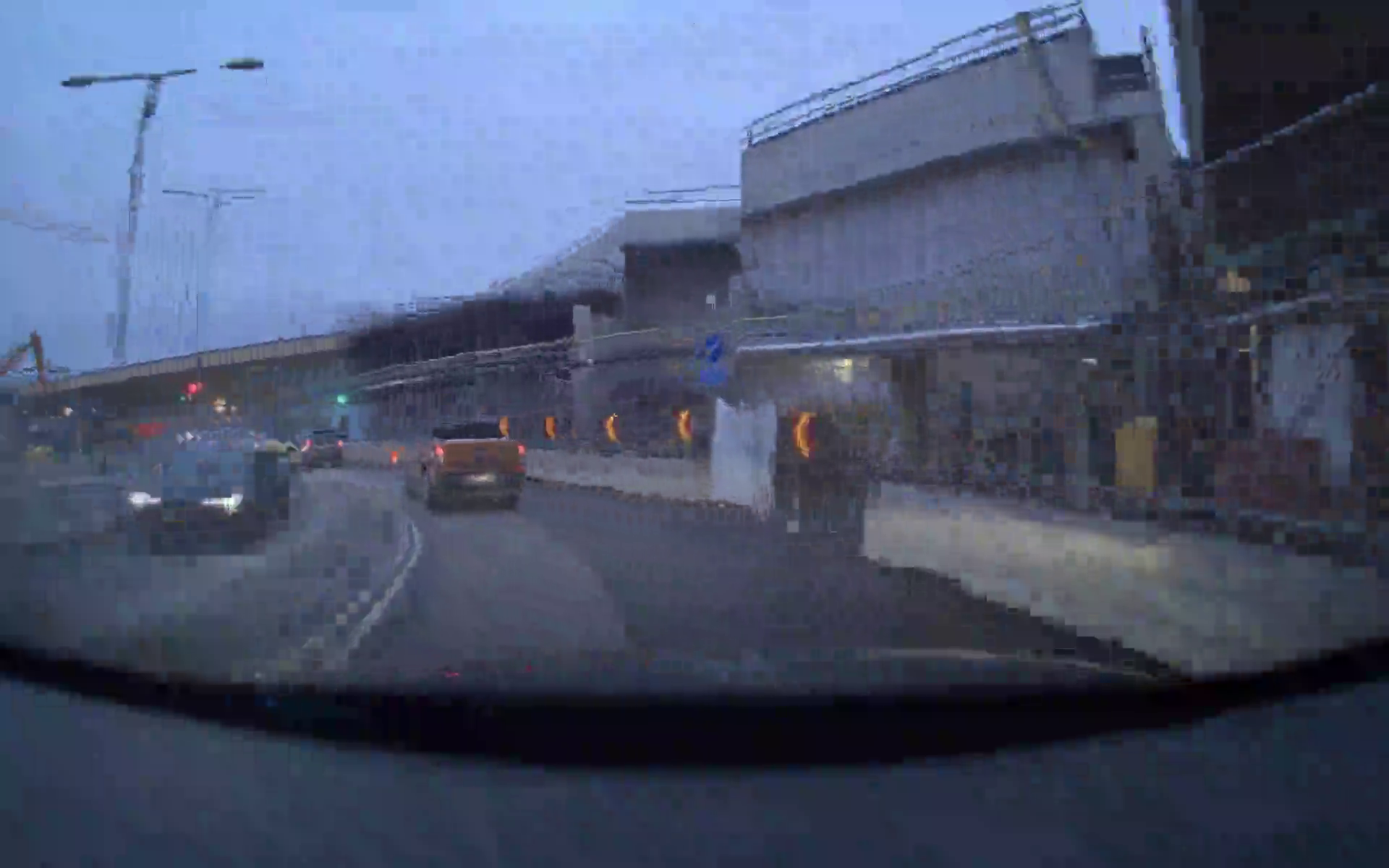}
    \caption{Compressed Video Negative Outlier (CRF 48)}
    \label{fig:negative}

    \includegraphics[width=\linewidth]{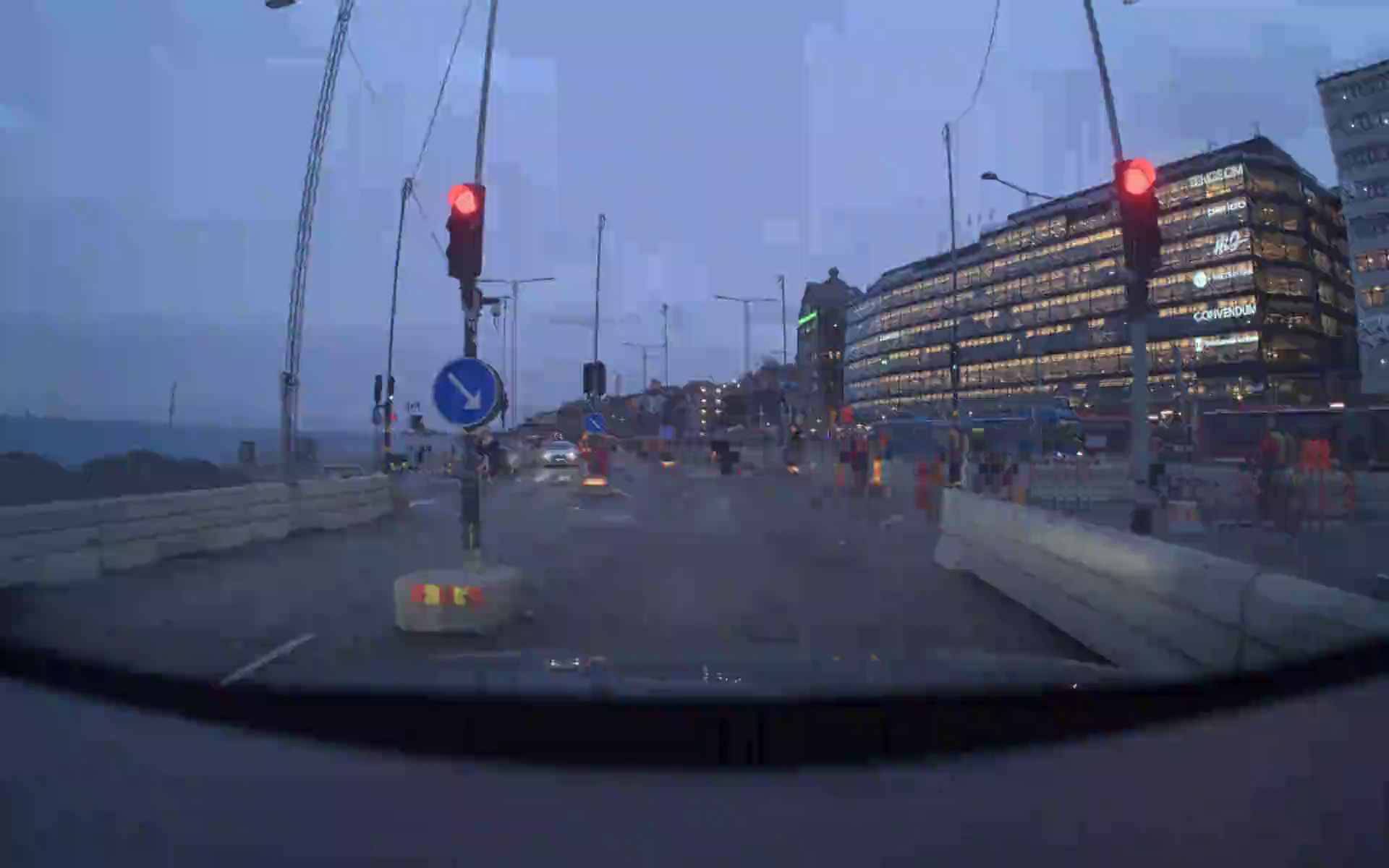}
    \caption{Compressed Video Positive Outlier (CRF 48)}
    \label{fig:positive}
 
    \caption{Examples of the largest negative and positive outliers between subjective assessment and VMAF prediction using the compressed video version only.
The largest negative outlier shows a substantial overestimation by the model, with a subjective mean score of 38.8 across all questionnaire dimensions compared to a VMAF prediction of 68.9 ($\Delta = -30.1$). In contrast, the largest positive outlier illustrates an underestimation by the model, where the subjective mean score was 55.8 while the VMAF prediction reached only 46.2 ($\Delta = +9.6$). These examples highlight systematic mismatches between model-based predictions and human perception for specific scenes.}

  \label{fig:compressed outliers}
\end{figure}

\subsubsection{Effects of Environmental Categories}

To examine potential contextual influences, the results were analyzed by weather condition (good vs. bad) and time of day (day vs. night).
Normality assumptions were violated in several category samples (Shapiro–Wilk, $p < .05$); hence, non-parametric statistics were applied.

A Kruskal–Wallis test across the three combined environmental categories (day–good, day–bad, night–good) showed no significant main effect ($H = 4.83$, $p = .089$).
Similarly, separate Welch’s t-tests revealed no significant differences between good and bad weather ($t = 1.38$, $p = .173$) or between day and night scenes ($t = 0.80$, $p = .426$).

Although the effects were not significant, a slight trend toward lower perceived quality in bad weather conditions was observed, particularly at higher compression levels (CRF~$\geq$~36).
Variance was marginally higher in night scenes, likely due to increased noise visibility and reduced contrast.
Overall, environmental context made only a minor contribution to perceived video quality differences, confirming that compression level remains the dominant determinant of perceived quality in teleoperation scenarios.

\subsection{Model–Human Alignment}

The improvement achieved by the newly trained model was measured by comparing the MOS values assigned by participants with the scores predicted by both models. 
Spearman’s $\rho$, Pearson’s $r$, and the Root Mean Square Error (RMSE) were used to quantify agreement between subjective and objective ratings. 

Across all scenes, the fine-tuned model achieved a Pearson correlation of $r = 0.87$ and a Spearman correlation of $\rho = 0.86$, matching the VMAF-4K baseline ($r = 0.88$, $\rho = 0.87$) in correlation strength but substantially reducing absolute error. 
The RMSE decreased from 10.36 to 8.83 and the MAD from 8.71 to 6.38, corresponding to improvements of approximately 15\% and 27\%, respectively. 
These results indicate that while both models track human judgments similarly in rank order, the fine-tuned variant yields markedly lower deviations from human ratings. 
The largest improvements were observed at medium compression levels (CRF~36–42), where the standard VMAF frequently underestimated the perceived degradation compared to human observers. 
A summary of these results is provided in Table~\ref{tab:model-results}.

\begin{table}[h]
  \centering
\small
  \caption{Correlation and error metrics between MOS and model predictions (overall).}
  \label{tab:model-results}
  \begin{tabular}{lrrrrr}
       \textbf{Model} & \textbf{MAD} & \textbf{MSE} & \textbf{RMSE} & \textbf{$r$} & \textbf{$\rho$} \\\\
    Finetuned Model & 4.94 & 43.37 & 6.59 & 0.910 & 0.899 \\
    VMAF-4K & 7.27 & 77.70 & 8.81 & 0.912 & 0.898 \\

  \end{tabular}
\end{table}

Qualitative inspection of the largest outliers (Figures~\ref{fig:compressed outliers} and~\ref{fig:positive}) revealed that discrepancies primarily occurred in scenes with strong reflections, rain droplets, or dynamic lighting. 
In these cases, the standard VMAF underestimated degradation, likely due to the feature weighting learned on entertainment content. 
The retrained model showed higher sensitivity to these effects, aligning more closely with subjective perception.

\section{Discussion}
The present findings highlight both the potential and the limitations of adapting existing video quality metrics to teleoperation. While the retrained model showed improved alignment with human judgments, the analysis also revealed critical boundary conditions where conventional feature weighting fails to capture task relevance.
\subsection{Questionnaire}
As can be seen in the results, the questionnaire shows both strong internal consistency and clear differentiation between the compression levels. This validates that participants were able to perceive and evaluate the deterioration in video quality with a high degree of reliability. 
Although the results suggest good construct validity in the experimental context, the ratings obtained merely reflect subjective impressions. However, these ratings cannot necessarily be interpreted as direct behavioral outcomes such as reaction time, braking accuracy, or time to collision.  It is possible that the subjects overestimate or underestimate themselves, especially since the majority have never remotely controlled a vehicle before.
Therefore, the questionnaire provides a good insight into perception, but it must be noted that this is not equivalent to an objective measure of performance. 
The proposed model can thus serve as an early indicator for identifying potentially critical quality deterioration that may impair operator performance. However, it is important to note the need for further empirical validation to establish quantitative relationships between perceived quality and safety-related variables.  
Future studies should explicitly combine subjective assessments with physiological and behavioral measures such as gaze dynamics, reaction latency, or driving precision to assess whether the deterioration in perception measured by VMAF or MOS actually correlates with operator safety and task performance.   
Consequently, the present instrument provides a solid basis for perception analysis, but its applicability to universal teleoperation systems  or concepts must be substantiated by additional validation studies.

\subsection{Fine-tuned Model}

The qualitative analysis of the largest deviations between MOS and the new VMAF model (Figures~\ref{fig:negative} and \ref{fig:positive}) provides deeper insights into the domain-specific challenges of video quality assessment in teleoperation. 
In the case of negative outliers, the model significantly overestimated the perceived quality. 
Although the text on the building facades remained legible, the middle area of the scene was heavily affected by compression artifacts, making it almost impossible to recognize the cyclists—objects that are crucial for the driving task. However, since this area only makes up a small part of the image, it carries little weight in the metric, as the rest, which is much larger, is rendered very well. This is because the cyclists are not static like the rest of the scene, and experience shows that motion is always difficult to compress.
This discrepancy shows that, from a teleoperation perspective, perception-relevant areas are not evenly distributed across the image and may only make up small areas, but are all the more important for that.
While text details are visible, the deterioration in areas important for the driving task (such as dynamic obstacles or other road users) has a major impact on perceived task performance and situational awareness, according to human evaluations.

This is also shown by the positive outlier in Figure \ref{fig:positive}, which contains a construction site with numerous details, such as cables or scaffolding, that were severely affected by the compression and many details were lost. 
Despite the loss of detail in the environment, the participants rated this sequence significantly more positively than the VMAF model. 
The reasons for this can be seen in the center line of the road and the outer boundary markings. These remained clearly visible, and their high visual conspicuity due to the neon orange color of the construction site markings and their size still allows for safe lane guidance. The participants prioritized these features, which are relevant to the driving task, over visual fidelity in peripheral areas that are not directly related to the driving task.
This once again underscores that perceived quality in teleoperation is context-dependent and primarily task-oriented. As has already been shown in studies on selective compression \cite{drorOptimizingTrafficSigns2024a, neumeierDataRateReduction2022}, artifacts or missing details in these areas have less of an impact on perceived quality in terms of teleoperation.

These results suggest that conventional video quality metrics such as VMAF or MS-SSIM, which weight all areas equally, are not entirely suitable for safety-critical applications such as teleoperation. 
These metrics assume that the entire image is equally important and prioritize based on the size of the artifacts rather than their position above important objects, whereas in teleoperation, only certain areas such as the navigable corridor, obstacles, and moving objects or people have the greatest semantic significance. 
The current model therefore still misjudges cases in which the deterioration affects task-relevant areas more than peripheral or less important areas.

\subsection{Limitations and future work}
A promising approach to overcoming this limitation is to extend a metric to include spatial weighting or semantic segmentation. 
Semantic weighting on the region of interest could prioritize areas with road markings, vehicles, or pedestrians, allowing the metric to better approximate human relevance assessment.
Another approach would be to improve the weightings of the VMAF model by training with larger and more diverse datasets. 
However, such datasets are difficult to generate, as the ITU~\cite{ITU_T_P910_2023} recommends at least 15 independent human evaluations per video sequence to ensure valid values. This requires suitable videos to be identified or recorded in advance.
This makes the creation of large-scale domain-specific datasets costly.

\section{Conclusion}

This study presented a domain-specific approach to evaluating video quality after compression in teleoperation, in which subjective ratings were used to create a VMAF model adapted to the new requirements. The results show that metrics such as VMAF can approximate perceived quality in general video content, but their predictive accuracy decreases in safety-critical, task-oriented contexts. By retraining the model with teleoperation-specific MOS, we achieved significantly better agreement with human perception, suggesting that context-sensitive adaptation can greatly improve the assessment of perceived video quality. The newly developed questionnaire proved effective in capturing subjective impressions regarding drivability, situational awareness, and perceptual clarity, as well as reflexes based on the original video. Although it does not capture objective data, it offers a valuable contribution to identifying conditions under which deteriorating visibility can impair long-distance driving performance. Future research should therefore combine subjective and objective measures to further validate the relationship between perception and performance. The qualitative analysis of outliers revealed that perceived video quality strongly depends on the semantic relevance of the impaired areas to the driving task. In teleoperation, artifacts and loss of detail in task-critical areas (e.g., road users or navigable space) carry greater weight than peripheral artifacts. This finding suggests that a domain-specific, context-sensitive extension of existing video image quality metrics should be considered, such as spatial weighting or segmentation-based learning, in order to better capture task relevance and enable optimal image quality assessment.
Overall, the proposed methodology and resulting model provide a basis for perception and function-oriented evaluation of video quality in teleoperation.

\section*{Ethical Considerations}

The user study was conducted in accordance with institutional and GDPR data-protection requirements. 
The study protocol was approved by the responsible non-medical ethics committee of an European technical university

\bibliographystyle{IEEEtran}
\bibliography{references}

\end{document}